\documentclass{aa}
\usepackage{lineno}
\usepackage[utf8]{inputenc}
\usepackage{natbib}
\usepackage{amsmath}    
\usepackage{subfigure}
\usepackage{graphicx}
\usepackage{float}
\usepackage{natbib}
\usepackage{dcolumn}
\usepackage{bm}
\usepackage{ulem}
\usepackage{color}
\usepackage{txfonts}
\usepackage{rotating}
\usepackage{booktabs}
\usepackage{array}
\usepackage{tabularx}
\usepackage{threeparttable}
\usepackage{diagbox}
\usepackage{multirow}
\usepackage{makecell}
\usepackage{longtable}
\usepackage{mhchem}
\usepackage{afterpage}
\usepackage{xcolor}
\usepackage{setspace}
\usepackage{geometry}
\usepackage[colorlinks=true, allcolors=blue]{hyperref} 
\definecolor{myblue}{RGB}{0,0,255}  
\definecolor{myblack}{RGB}{0,0,0}   

\begin{document}
\nolinenumbers

\title{Formation of the Dormant Black Holes with Luminous Companions from Binary or Triple Systems}

\author{Zhuowen Li\inst{1}
\and    Xizhen Lu\inst{1}
\and    Guoliang L\"{u}\inst{1,2*}
\and    Chunhua Zhu\inst{1}
\and    Helei Liu\inst{1}
\and    Li Lei\inst{1}
\and    Sufen Guo\inst{1}
\and    Xiaolong He\inst{1,2}
\and    Nurzada Beissen\inst{3}
}
 \institute{School of Physical Science and Technology,Xinjiang University, Urumqi, 830046, China\\
              \email{guolianglv@sina.com, chunhuazhu@sina.cn, guosufen@xju.edu.cn}
         \and Xinjiang Astronomical Observatory, Chinese Academy of Sciences, 150 Science 1-Street, Urumqi, Xinjiang 830011, China
         \and Institute for Experimental and Theoretical Physics, Al-Farabi Kazakh National University, Almaty 050040, Kazakhstan}

  \abstract
   {Recently, a class of dormant black hole binaries with luminous companions (dBH-LC) has been observed, such as $Gaia$ BH1, BH2, and BH3. Unlike previously discovered X-ray BH binaries, this type of dBH-LC has relatively long orbital periods (typically more than several tens to a few hundred days) and shows very weak X-ray emission. Therefore, studying the formation and evolution of the whole dBH-LC population is also a very interesting problem.}
   {Our aim is to study the contribution of massive stars to the dBH-LC population under different evolutionary models (isolated binary evolution (IBE) and hierarchical triple evolution), and different formation channels (such as mass transfer, common envelope evolution).}
   {Using the Massive Objects in Binary Stellar Evolution (MOBSE) code, the Triple Stellar Evolution (TSE) code, and the latest initial multiple-star distributions, we model the populations of massive stars. Finally, we calculate the orbital properties, mass distributions, and birthrates of the BH-LC populations formed under these different conditions.}
   {In the Milky Way, we calculate that the birthrate of dBH-LC formed through IBE is about 4.35$\times$$10^{-5}$ ${\rm yr}^{-1}$, while the birthrate through triple evolution is about 1.47$\times$$10^{-3}$ ${\rm yr}^{-1}$. This means that the birthrate from triple evolution is one to two orders of magnitude higher than that from IBE. We find that in triple evolution, the main formation channel of dBH-LC is post-merger binaries formed from inner binary mergers triggered by von Zeipel$-$Lidov$-$Kozai oscillations. In particular, if the merger product is formed from a centre-helium-burning star and a main-sequence star, the resulting star usually has a small core mass and a large envelope mass. These stars with this structure can form BHs in the pair-instability supernova (PISN) range (about 60 $M_{\odot}$$\sim$120 $M_{\odot}$), which are about 3 times more massive than the maximum BH mass formed through IBE.}
   {Due to the presence of dynamical effects in triple evolution, the inner binaries in triples are more likely to interact than isolated binaries. As a result, dBH-LC formed through triple evolution mainly come from the channel where the inner binary merges. The birthrate of dBH-LC from triple evolution is about two orders of magnitude higher than that from IBE. In addition, some dBH-LC with heavy BHs are also formed through the inner binary merger channel in triple. These results strongly indicate that the triple evolution can be the most important channel for dBH-LC formation.}


   \maketitle

\section{Introduction}
It is estimated that the Milky Way (MW) contains about $10^{8}$ stellar-mass black holes (BHs) \citep{1994ApJ...423..659B,1996ApJ...457..834T}. These BHs may exist in single-star systems, binary systems, or multiple-star systems \citep{2006csxs.book..157M,2006ARA&A..44...49R,2008ApJ...687..466Q,2020A&A...641A..43B,2021MNRAS.507.2804W,2025ApJ...983..104S}. In the past, BHs with luminous companions (LCs) were usually discovered using radio and X-ray observations \citep{2006csxs.book..157M,2006ARA&A..44...49R}. In the catalog by \cite{2016A&A...587A..61C}, 59 X-ray binaries containing BHs were confirmed, they also estimated the masses and orbital properties of these BH X-ray sources. Recent breakthroughs in radial velocity measurements and astrometric techniques have revealed a population of dormant black hole with LC systems (dBH-LCs) characterized by low X-ray luminosity, wide orbital separations, and negligible mass transfer  \citep{2019Natur.575..618L,2022arXiv220700680A,2023AJ....166....6C,2023ApJ...946...79T,2023MNRAS.518.1057E,2023MNRAS.521.4323E,2024A&A...686L...2G,2024NatAs...8.1583W}. Table 1 shows the physical properties of some recently discovered typical dBH-LCs. These dBH-LCs exhibit distinctive wide orbits and extreme mass ratios, making them pristine laboratories for testing binary and multiple-star evolution theories.

Most population synthesis studies of dBH-LCs focus on the isolated binary evolution (IBE) model, which predicts that the birthrate of dBH-LCs is in the range of $\sim$$10^{-5}$ ${\rm yr}^{-1}$ to $\sim$$10^{-4}$ ${\rm yr}^{-1}$ \citep{2019ApJ...885..151S}, or their number is between a few hundred and several tens of thousands \citep{2017ApJ...850L..13B,2018ApJ...861...21Y,2022ApJ...931..107C,2024ApJ...975..163C}. However, the orbital period distribution of dBH-LCs formed by the IBE model shows a bimodal shape, with a gap between $10^{2}$ days and $10^{4}$ days \citep{2025PASP..137d4202N}. This makes it difficult for the IBE model to explain the orbital properties of some recently observed dBH-LCs, such as Gaia BH1 and BH2 \citep{2023MNRAS.518.1057E,2023MNRAS.521.4323E}. On the other hand, the dynamical model in globular cluster has also been proposed as an alternative way to form dBH-LCs \citep{2023MNRAS.526..740R,2024A&A...688L...2M,2024ApJ...965...22D}. In the simulations by \cite{2024ApJ...965...22D}, the formation efficiency of dBH-LCs in globular clusters (GCs) is about 50 times higher than that in the IBE model. Later, the work by \cite{2025PASP..137d4202N} further compared the physical properties and formation efficiencies of dBH-LCs formed in GCs and through IBE. However, the dynamical model is generally only applicable in environments with relatively high stellar densities.

Considering that about 73\%$\pm$16\% of massive stars are in triple or higher-order systems \citep{2017ApJS..230...15M}, the IBE model may not be the best explanation for the formation of dBH-LCs. In hierarchical triples, long-term ZLK \citep{1910AN....183..345V,1962P&SS....9..719L,1962AJ.....67..591K,2016ARA&A..54..441N} effects during evolution often cause oscillations in the eccentricity of the inner binary (which refers to the two stars in the closer orbit). Therefore, compared to the IBE model, the evolution of massive stars in triple systems is more likely to lead to interactions at periastron \citep{2009ApJ...697.1048P,2022MNRAS.515L..50V}. In addition, triple evolution can produce many novel evolutionary pathways. Several typical observational phenomena can be explained by triple evolution \citep{2000ApJ...535..385F,2003ApJ...598..419W,2014ApJ...781...45A,2014ApJ...793..137N,2016MNRAS.460.3494S,2016ApJ...816...65A,2016ApJ...822L..24N,2017ApJ...841...77A,2018A&A...610A..22T,
2019MNRAS.487.3029S,2019ApJ...881...41L,2020ApJ...903...67M,2020A&A...640A..16T,2020ApJ...903...67M,2020ApJ...900...16F,2020ApJ...895L..15F,2020ApJ...903...67M,2023arXiv230210350B,2023ApJ...950....9R,2023A&A...678A..60K,
2023PhRvD.107d3009X,2023ApJ...955L..14S,2024ApJ...975L...8L,2025ApJ...979L..37L,2025ApJ...983..115S,2025A&A...693A..14B,2025ApJ...978...47S,2025ApJ...991L..54S,2025A&A...698A.240S,2025PASP..137g4201S,2025A&A...699A.272V,2025arXiv250913009K,2025arXiv250813264X,2025ApJ...985L..42X}.
Most previous studies on the dBH-LC population focused on the IBE model\citep{2017ApJ...850L..13B,2019ApJ...885..151S,2022ApJ...931..107C,2024A&A...690A.144I,2024ApJ...975..163C},  and some dBH-LC systems formed through triple evolution mainly focused on observed like-Gaia systems (such as Gaia BH1/2, and G3425) \citep{2024ApJ...975L...8L,2024ApJ...964...83G,2025ApJ...979L..37L,2025ApJ...988L...7R,2025ApJ...992L..12N}. However, there are still few studies on the contribution of triple evolution to the entire dBH-LC population. Therefore, in this work, we study whether triple evolution can form dBH-LCs, in order to evaluate its importance to the entire dBH-LC population.

In this work, we perform the formation and evolution of dBH-LCs from binary or triple systems via population synthesis method. We consider the evolution of massive stars in isolated binary and hierarchical triple, to predict the parameter properties and birthrate of dBH-LCs. In Section 2, we introduce the methods for modeling the evolution of massive stars in isolated binary and hierarchical triple, and the approach used for population synthesis calculations. In Section 3, we present the calculation results of dBH-LCs. In Sections 4 and 5, we discuss and summarize the results.

\begin{table*}[ht]
\centering

\caption{The physical parameters of dBH-LCs observed through radial velocity or astrometric measurements are listed.}
\label{tab:1}

\begin{threeparttable}

\begin{tabular}{lcccccc}
\hline\hline
Name & $M_{\mathrm{BH}}$ ($M_\odot$) & $M_{\mathrm{LC}}$ ($M_\odot$) & $P_{\mathrm{orb}}$ (days) & $e$ & Type \\
\hline
GC NGC3201 \#12560 & $\geq 4.36$ & $0.81^{+0.05}_{-0.05}$ & $166.88^{+0.71}_{-0.63}$ & $0.610^{+0.020}_{-0.020}$ & MS \\
GC NGC3201 \#21859 & $\geq 7.68$ & $0.61^{+0.05}_{-0.05}$ & $2.24^{+0.01}_{-0.01}$ & $0.070^{+0.04}_{-0.04}$ & MS \\
VFTS 243 & $\geq 8.70$ & $25.0^{+2.3}_{-2.3}$ & $10.40^{+0.01}_{-0.01}$ & $0.017^{+0.01}_{-0.01}$ & MS \\
HD 130298 & $8.80^{+3.5}_{-1.5}$ & $24.2^{+3.8}_{-3.8}$ & $14.63^{+0.01}_{-0.01}$ & $0.457^{+0.007}_{-0.007}$ & MS \\
Gaia BH1 & $9.78^{+0.18}_{-0.18}$ & $0.93^{+0.05}_{-0.05}$ & $185.59^{+0.05}_{-0.05}$ & $0.454^{+0.005}_{-0.005}$ & MS \\
AS 386 & $\geq 7.00$ & $7^{+1}_{-1}$ & $131.27^{+0.09}_{-0.09}$ & $0$ & Giant \\
Gaia BH2 & $8.93^{+0.33}_{-0.33}$ & $1.07^{+0.19}_{-0.19}$ & $1276.70^{+0.6}_{-0.6}$ & $0.518^{+0.002}_{-0.002}$ & Giant \\
Gaia BH3 & $32.70^{+0.82}_{-0.82}$ & $0.76^{+0.05}_{-0.05}$ & $4253.1^{+98.5}_{-98.5}$ & $0.7291^{+0.0048}_{-0.0048}$ & Giant \\
2M05215658+4359220 & $3.30^{+0.80}_{-0.70}$ & $4.4^{+2.2}_{-1.5}$ & $82.20^{+2.50}_{-2.50}$ & $0.005^{+0.003}_{-0.003}$ & Giant \\
Gaia ID 3425577610762832384 & $3.60^{+0.80}_{-0.50}$ & $2.66^{+1.18}_{-0.68}$ & $877^{+2.0}_{-2.0}$ & $0.05^{+0.01}_{-0.01}$ & Giant \\
\hline
\end{tabular}

\begin{tablenotes}
\footnotesize
\item \textbf{Notes.}
The second and third columns show the mass of BH ($M_{\rm BH}$) and the mass of LC ($M_{\rm LC}$), respectively. The fourth and fifth columns give the orbital period ($P_{\rm orb}$) and eccentricity ($e$) of the dBH-LCs. The last column indicates the type of LC. The data from the first to the last row come from sources \cite{2018MNRAS.475L..15G}, \cite{2019A&A...632A...3G}, \cite{2022NatAs...6.1085S}, \cite{2022A&A...664A.159M}, \cite{2023MNRAS.518.1057E}, \cite{2018ApJ...856..158K}, \cite{2023MNRAS.521.4323E}, \cite{2024A&A...686L...2G}, \cite{2019Sci...366..637T}, and \cite{2024NatAs...8.1583W}, respectively. In addition, we refer to the recent review on BH by \cite{2023arXiv231115778C} (and the references within). Some observations of dBH-LC that are still under debate are not included in our statistics (for example, the findings by \cite{2014Natur.505..378C}, \cite{2019Natur.575..618L}, \cite{2019Sci...366..637T}, \cite{2020A&A...637L...3R}, \cite{2021MNRAS.504.2577J}, \cite{2022MNRAS.511.2914S} and \cite{2022A&A...665A.180L}).).
\end{tablenotes}

\end{threeparttable}
\end{table*}

\section{Methodology}
For the dBH-LCs, BH accretes almost no material, so their X-ray emission can be neglected. We follow the method of \cite{2024A&A...690A.256S} to identify dBH-LCs with very low X-ray luminosity ($L_{\rm X}$). Specifically, we first use Equation 20 from \cite{2024A&A...690A.256S} to determine the critical condition for the formation of an accretion disc, given by
\begin{equation}
P_{\mathrm{orb}} < 4\pi G M_{\mathrm{BH}} c \left( \frac{\eta}{f(\bar{s})} \right) \frac{1}{v_{\mathrm{w}}^4}
\label{eq:critical_orbital_period}
\end{equation}
where $G$ is the gravitational constant and $c$ is the speed of light, $v_{\rm w}$ is the wind speed. We use the method of \cite{2008ApJS..174..223B} to calculate $v_{\rm w}$, that is,
\begin{equation}
V_{\text{w}}^{2} = 2 \beta_{\text{wind}} \frac{G M_{\text{LC}}}{R_{\text{LC}}} \, ,
\end{equation}
here, $\beta_{\rm wind}$ is related to the spectral type of the LC, that is, \begin{equation}
\beta_{\rm wind} =
\begin{cases}
0.125, & \text{giants} \\[6pt]
7, & \text{MS and } M_{\text{MS}} > 120\,M_{\odot} \\[6pt]
0.5, & \text{MS and } M_{\text{MS}} < 1.4\,M_{\odot} \\[6pt]
7, & \text{He stars and } M_{\text{He}} > 120\,M_{\odot} \\[6pt]
0.125, & \text{He stars and } M_{\text{He}} < 10\,M_{\odot}
\end{cases}
\end{equation}
For MS stars or He stars with masses between the two extreme values, $\beta_{\rm wind}$ is calculated by linear interpolation. The constants $\eta$ and ${f(\bar{s})}$ come from hydrodynamical simulations and are usually taken as $\eta = \frac{1}{3}$ and ${f(\bar{s})}$ = $\sqrt{12}$. When a dBH-LC forms an accretion disc, we calculate $L_{\rm X}$ using Equations (21) $\sim$ (30) in Section 3.3 of \cite{2024A&A...690A.256S}. Otherwise, we use Equations (31)$\sim$(32) in Section 3.4 to compute $L_{\rm X}$. In a dBH-LC, the BH usually does not show clear accretion (typically with an accretion rate lower than $10^{-14}$ M$_\odot$/ ${\rm yr}$ $\sim$$10^{-13}$ M$_\odot$/ ${\rm yr}$, see Figure 1 in \cite{2024A&A...690A.256S}). Observationally, the X-ray flux (or $L_{\rm X}$) of a dBH-LC is very low. For example, the $L_{\rm X}$ of Gaia BH1/2 are about
$10^{29.4}$ erg/s and $10^{30.1}$ erg/s, respectively \citep{2023arXiv231105685R}. In addition, \cite{2020A&A...636A..99V} pointed out that $L_{\rm X} \sim 10^{35}$ erg/s is roughly the detection threshold of all-sky X-ray instruments. The Chandra T-ReX programme \citep{2022MNRAS.515.4130C} also shows that BH binaries with $L_{\rm X}$ between $10^{31}$ erg/s and $10^{35}$ erg/s can be classified as dBH-LCs. Therefore, following the approach of \cite{2020ApJ...898..143S} and \cite{2024A&A...690A.256S}, we define a dBH-LC as dormant and detached if $L_{\rm X} < 10^{35}$ ${\rm erg/s}$ and it does not fill its Roche lobe.

In the following subsections, we introduce the stellar evolution codes used to model the formation of dBH-LC populations from massive stars through IBE and triple evolution. We also describe the initial parameter distributions and the calculation of birthrates.

\subsection{Binary evolution}
We use the Binary Stellar Evolution (BSE) code originally developed by \cite{2000MNRAS.315..543H,2002MNRAS.329..897H}, along with the updated version for massive star evolution, MOBSE (which stands for ‘Massive Objects in Binary Stellar Evolution), which was revised by \cite{2018MNRAS.474.2959G}. MOBSE includes updated prescriptions for stellar winds, SN kicks, and BH formation (including the effects of pair-instability SN (PISN) \citep{1983A&A...119...61O,1984ApJ...280..825B,2003ApJ...591..288H,2007Natur.450..390W} and pulsational pair-instability SN (PPISN) \citep{1967PhRvL..18..379B,2007Natur.450..390W,2014ApJ...792...28C,2016PhRvD..93l3012Y}.

For the initial rotation velocity of massive stars, we sample from the empirical distribution derived by \cite{2013A&A...560A..29R} based on observations of 216 O-type stars. For the remnant mass prescription, we adopt the "delayed" model from \cite{2012ApJ...749...91F}. For the SN kick model, we use the prescription from \cite{2020ApJ...891..141G}, where the kick velocity is given by
\begin{equation}
v_{\mathrm{kick}} = f_{\mathrm{H05}} \left( \frac{\langle M_{\mathrm{NS}} \rangle}{M_{\mathrm{rem}}} \right) \left( \frac{M_{\mathrm{ej}}}{\langle M_{\mathrm{ej}} \rangle} \right)
\end{equation}
Here, $f_{\mathrm{H05}}$ is a random number drawn from a Maxwellian distribution with a one-dimensional rms $\sigma = 265\ \mathrm{km\ s^{-1}}$ \citep{2005MNRAS.360..974H}. $\langle M_{\mathrm{NS}} \rangle$ represents the average mass of neutron stars, and $\langle M_{\mathrm{ej}} \rangle$ is the average ejected mass associated with the formation of a neutron star of mass $\langle M_{\mathrm{NS}} \rangle$ through single star evolution.

For stable MT phases, we assume non-conservative, with a fixed accretion efficiency of $\beta$ = 0.5 \citep{2020MNRAS.498.4705V,2025A&A...696A..54S}. In addition, we assume that the companion can survive a common envelope (CE) phase initiated by a Hertzsprung gap donor. MOBSE treats the CE phase using the standard $\alpha_{\rm CE}$-$\lambda$ formalism, where $\alpha_{\rm CE}$ represents the fraction of orbital energy used to unbind the envelope, and $\lambda$ is the structural parameter of the envelope. We refer to recent 3D CE simulations by \cite{2024A&A...691A.244V}, in which the value of $\alpha_{\rm CE}$ typically ranges from 0.5 to 2.57. Thus, we adopt the intermediate value $\alpha_{\rm CE} = 1.5$ as the default. The value of $\lambda$ is calculated following appendix A of \cite{2014A&A...563A..83C}. For merger products formed through CE or contact, we adopt the default prescription of MOBSE to calculate their structure, mass, age, spin, stellar type, and so on. When considering that the later evolution of some merger products may contribute to the dBH-LC population, we mainly distinguish the following three scenarios: 1. When two stars without a clear helium core (i.e., MS stars) merge, the merger product is still an MS star. 2. When a star with a clear helium core (such as an Hertzsprung Gap (HG) star, a center-helium-burning (CHeB) star, or a giant) merges with an MS star, the merger product is a star with a helium core. Its helium core mass remains unchanged, while the mass of the MS star is absorbed into the envelope of the helium-core star. 3. When two stars with clear helium cores merge, the merger product is still a star with a helium core. The helium core mass of the product is equal to the sum of the helium core masses of the two parent stars.

In addition, the MOBSE code is based on the fitting formulae by \cite{2000MNRAS.315..543H,2002MNRAS.329..897H} (the SSE model). Its main advantage is computational efficiency, and it has been widely used in studies of the dBH-LC population. However, recent studies by \cite{2025A&A...698A.240S} and \cite{2025ApJ...983..115S} have pointed out that the SSE model often overestimates the maximum radius of stars. For more discussion on the differences between the SSE model and detailed stellar evolution codes, see Section 4.1.

\subsection{Triple stellar evolution}
We use the triple stellar evolution (TSE) code for massive stars developed by \cite{2022MNRAS.516.1406S} and \cite{2024JOSS....9.7102S} to study the formation of dBH-LC populations through the triple evolution channel. The advantage of this code is that it incorporates up-to-date prescriptions for massive star evolution (i.e., those from MOBSE) and post-Newtonian (PN) terms up to the 2.5PN order. In addition, TSE self-consistently couples stellar evolution, binary interactions (such as tides and MT) with long-term three-body dynamical equations, including terms up to the octupole-level approximation \citep{2013MNRAS.431.2155N}. Specifically, TSE first uses the single-star evolution from MOBSE to compute the masses, radii, and other properties of the three stars as functions of time. It then integrates the long-term gravitational dynamics of the triple system using Equations (6) to (9) from \cite{2022MNRAS.516.1406S} to calculate the time evolution of the orbital parameters and stellar spins. During the integration process, if the inner binary undergoes Roche-lobe overflow (RLOF), TSE passes the corresponding stellar and orbital parameters to MOBSE to simulate MT phase (For more details on TSE, see Section 2.4 of \cite{2022MNRAS.516.1406S}. Therefore, all modifications described in Section 2.1 also apply to the TSE code. In addition, we use the default tidal model from the work of \cite{2022MNRAS.516.1406S}. Specifically, in this model, the equilibrium tide equations of \cite{2011CeMDA.111..105C,2016CeMDA.126..189C} are used for all types of stars. These equations include both dissipative and non-dissipative terms (see Appendix A of \cite{2018MNRAS.479.4749B}). The time derivatives of the stellar spin angular momentum ($\dot{L_{\rm in}}$), the eccentricity ($\dot{e_{\rm in}}$), and the semi-major axis ($\dot{a_{\rm in}}$) of the inner binary due to tidal effects can be described by the following equations: \begin{equation}
\left\{
\begin{aligned}
\dot{L}_{\mathrm{in}} &\propto \sum_{i=1,2} \left[
\frac{ f_5(e_{\mathrm{in}}) \, \boldsymbol{\Omega}_i \cdot \hat{\boldsymbol{\jmath}}_{\mathrm{in}} }{ j_{\mathrm{in}}^9 \, \omega_{\mathrm{in}} }
- \frac{ f_2(e_{\mathrm{in}}) }{ j_{\mathrm{in}}^{12} }
\right] , \\
\frac{\dot{e}_{\mathrm{in}}}{e_{\mathrm{in}}} &\propto \sum_{i=1,2} \left[
\frac{11}{18} \cdot \frac{ f_4(e_{\mathrm{in}}) \, \boldsymbol{\Omega}_i \cdot \hat{\boldsymbol{\jmath}}_{\mathrm{in}} }{ j_{\mathrm{in}}^{10} \, \omega_{\mathrm{in}} }
- \frac{ f_3(e_{\mathrm{in}}) }{ j_{\mathrm{in}}^{13} }
\right] , \\
\frac{\dot{a}_{\mathrm{in}}}{a_{\mathrm{in}}} &\propto \sum_{i=1,2} \left[
\frac{ f_2(e_{\mathrm{in}}) \, \boldsymbol{\Omega}_i \cdot \hat{\boldsymbol{\jmath}}_{\mathrm{in}} }{ j_{\mathrm{in}}^{12} \, \omega_{\mathrm{in}} }
- \frac{ f_1(e_{\mathrm{in}}) }{ j_{\mathrm{in}}^{15} }
\right] .
\end{aligned}
\right.
\end{equation}
here, the polynomial function $f_{1,2,3,4,5}(e_{\rm in})$ is given in Appendix A of \cite{2022MNRAS.516.1406S}. The vector $\boldsymbol{\Omega_{i}}$ represents the stellar spin angular velocity. $j_{\rm in}$ and $\hat{\boldsymbol{\jmath}}_{\mathrm{in}}$ denote the dimensionless orbital angular momentum and its unit vector, respectively. $\omega_{\rm in}$ is the mean motion of the inner orbit, and the subscript $i$ refers to the two stars in the inner binary. This tidal model depends on two key parameters: the lag time $\Delta t$ and the second Love number $k_{\rm 2i}$. Following the assumptions in \cite{2011CeMDA.111..105C}, \cite{2016MNRAS.456.3671A} and \cite{2022MNRAS.516.1406S}, we set $\Delta t = 1\,\mathrm{s}$ and $k_{\rm 2i} = 0.028$.

Before the final integration time(which we set to the Hubble time, i.e., 13.7 Gyr), the TSE code stops the simulation if any of the following events occur:

(1) the tertiary fills its Roche lobe;

(2) the triple becomes dynamically unstable, when \citep{2001MNRAS.321..398M}
\begin{equation}
\frac{a_{\mathrm{out}}(1 - e_{\mathrm{out}})}{a_{\mathrm{in}}}
< 2.8 \left[ \left(1 + \frac{m_3}{m_{1}+m_{2}} \right)
\frac{1 + e_{\mathrm{out}}}{\sqrt{1 - e_{\mathrm{out}}}} \right]^{2/5}
\end{equation}
(Here, $a$ is the semi-major axis. The subscripts "in" and "out" refer to the inner and outer orbits, respectively, while the subscripts 1, 2, and 3 correspond to the three stars in the triple), at which point the secular approximation breaks down;

(3) the inner orbit is disrupted due to a SN.

It is worth noting that when the inner binary merges, we pass the properties of the merger product, the properties of the tertiary, and the outer orbital parameters as a new binary (the post-merger binary) to MOBSE for further evolution. This is slightly different from the termination condition used in \cite{2022MNRAS.516.1406S}, but it is consistent with that used by \cite{2023ApJ...955L..14S} and \cite{2025ApJ...978...47S}.

Considering the high computational cost of triple simulations, we set the maximum integration time for each triple to 5 hours. Less than 0.4\% of the sampled systems exceeded this limit. These systems are typically compact but dynamically stable triples with very short outer orbital periods. However, we do not expect this to significantly affect our results. In addition, due to computational limitations, this study is restricted to a single triple population. We did not explore the dependence of our results on model assumptions such as different supernova models, metallicities, or the CE parameter $\alpha_{\rm CE}$.

\subsection{Initial conditions}
Similar to \cite{2017ApJ...850L..13B} and \cite{2021ApJ...920...81S}, we assume a constant star formation rate (SFR) of $\mathrm{SFR}_{\mathrm{MW}} = 2.15\,M_\odot/\mathrm{yr}$ \citep{2003A&A...409..523R} over the past 13.7 Gyr for the Milky Way, and a metallicity of $Z = 0.014$ \citep{2013pss5.book..393Y}. This value is based on the recent observational results from \cite{2011AJ....142..197C}, who reported a Milky Way star formation rate of approximately $1.9\pm0.4\ M_{\odot}$.

For the initial distribution of binary, \cite{2017ApJS..230...15M} used spectroscopy, eclipses, long-baseline interferometry, adaptive optics, and common proper motion to study early-type binary observations. They found that the primary mass ($M_{\rm 1}$), the mass ratio ($q$), the $P_{\rm orb}$, and the $e$ are not independent, but show significant correlations. For the initial distribution of triple, \cite{2025PASP..137i4201S} used Gaia data to build $\sim$ 10,000 resolved triples within 500 pc of the Sun. From this, they obtained the distribution features of the intrinsic demographics of the triple population. Therefore, combining the statistical results of \cite{2017ApJS..230...15M} and \cite{2025PASP..137i4201S}, we use the following steps to sample the initial parameter distributions of binary and triple:

1. We first sample $M_{\rm 1}$ from the initial mass function of \cite{2001MNRAS.322..231K}. Considering the validity of dBH-LCs calculations, at a metallicity of Z = 0.014, we assume that the minimum initial stellar mass required to form a BH is $\sim$ 18 $M_{\odot}$. Therefore, in the sampling of binary and triple, we require $M_{\rm 1} > 18 M_{\odot}$ for binaries and ($M_{1} + M_{2}$) $>$ 18 $M_{\odot}$ for triples. In addition, the distributions by \cite{2017ApJS..230...15M} only give constraints for binaries with mass ratios $q \ge 0.1$. This is mainly because high brightness contrast makes astrometric, spectroscopic, and interferometric measurements difficult \citep{2023Galax..11...98M}. However, based on the parameters of some recently discovered dBH-LC systems, their progenitor mass ratios were much lower than 0.1. For example, $Gaia$ BH1, $Gaia$ BH2, and $Gaia$ BH3 all had initial mass ratios $q < 0.02$ \citep{2023Galax..11...98M,2024ApJ...964...83G,2024A&A...692A.141K,2024A&A...690A.144I}. Therefore, following \cite{2024ApJ...964...83G}, we extrapolate the mass ratio distribution given by \cite{2017ApJS..230...15M}.

2. Then, we use the method in Section 2.1 of \cite{2024ApJ...964...83G} to generate several companion periods from the companion frequency distribution function of this $M_{\rm 1}$ (we only allow at most two companion periods to be generated). This companion frequency distribution comes from the statistical results of \cite{2017ApJS..230...15M}. Through this simulation, we can determine whether this $M_{\rm 1}$ belongs to a single, a binary, or a triple. If this $M_{\rm 1}$ is born in a binary or triple, then this binary or this inner binary in the triple follows the distribution of \cite{2017ApJS..230...15M} \citep{2025PASP..137i4201S}.

3. For triples, following \cite{2025PASP..137i4201S}, we sample the outer mass ratio ($q_{\rm out} = M_{3} / (M_{1} + M_{2})$) from a power law with a logarithmic slope of $\gamma$ = –1.4. The outer orbital period is sampled from the log-normal distribution of \cite{1991A&A...248..485D}. The outer eccentricity is sampled from a thermal distribution. During the sampling process, we require that the initial triples to be hierarchical and satisfy the stability criterion. The stability criterion is given in Equation (5), and the condition for hierarchy is: \citep{2016ARA&A..54..441N}
\begin{equation}
\frac{a_{\rm in}}{a_{\rm out}} \cdot \frac{e_{\rm out}}{1 - e_{\rm out}^2} < 0.1 .
\end{equation}

\begin{figure*}[htb]
\centering
\includegraphics[height=9cm, keepaspectratio]{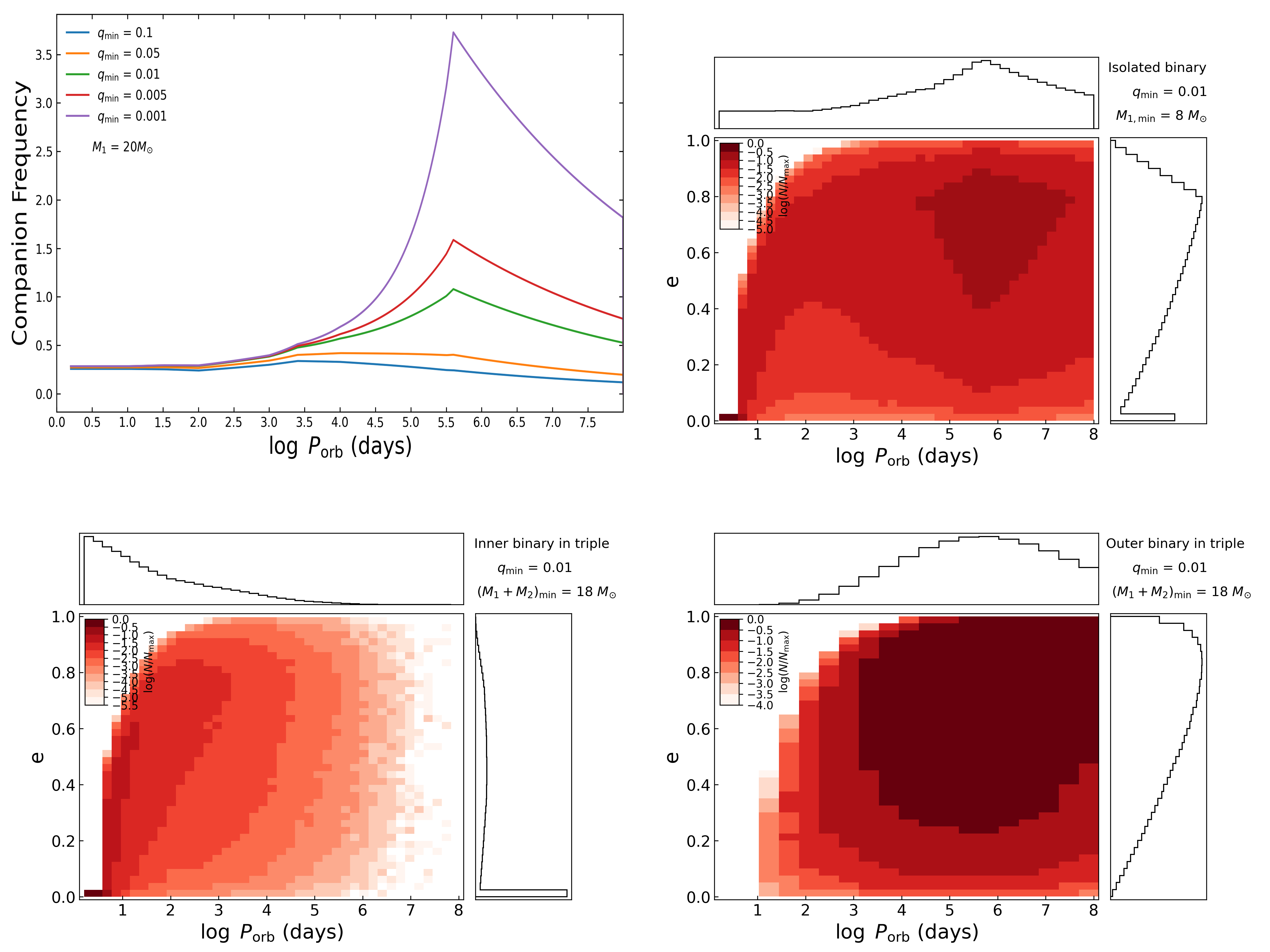}
\caption{The top-left panel shows the companion frequency for a primary mass of 20 $M_{\odot}$ at different orbital periods. The blue line represents the companion frequency from \cite{2017ApJS..230...15M} for a 20 $M_{\odot}$ primary, while the other colors show extrapolated companion frequencies for different minimum mass ratios ($q_{\rm min}$). The top-right panel shows the initial orbital distribution of isolated binaries with a $q_{\rm min}$ = 0.001. The two bottom panels show the initial inner and outer orbital distributions of triple with $q_{\rm min}$ = 0.001.}
\label{fig:1}
\end{figure*}

As shown in Fig. \ref{fig:1}, the mass ratio extrapolation affects the distribution of companion frequency with orbital period. As the minimum mass ratio is extended from 0.1 to 0.001, the peak of the companion frequency shifts from 3.5 to 5.5. In addition, the initial orbital parameter distributions of isolated binaries and triples are shown in Fig. \ref{fig:1}.

\subsection{Birthrate calculation}
We use a method similar to that of \cite{2016ApJ...822L..24N}, \cite{2025ApJ...983..115S} and \cite{2024ApJ...964...83G} to estimate the birthrate ($R$) of the dBH-LC population. Using the sampling described in Section 2.3, we perform $4 \times 10^5$ triple simulations and $4 \times 10^6$ binary simulations to reduce the uncertainty in the $R_{\text{birthrate}}$ calculation. Specifically, we use the following equation:
\begin{equation}
\left\{
\begin{aligned}
R_{\text{binary}} &= \frac{\text{SFR}_{\text{MW}}}{M_*}
\times f_{M_1 \geq 18 M_{\odot}}
\times f_{\text{binary}}
\times f_{\text{dBH-LC}} \\[6pt]
R_{\text{triple}} &= \frac{\text{SFR}_{\text{MW}}}{M_*}
\times f_{M_1 \geq 8 M_{\odot}}
\times f_{(M_1+M_2) \geq 18 M_{\odot}}
\times f_{\text{triple}}
\times f_{\text{dBH-LC}}
\end{aligned}
\right.
\end{equation}
where, $M_{*}$ is the average stellar mass calculated from the initial mass function of \cite{2001MNRAS.322..231K}, with a value of $\sim$ 0.572 $M_{\odot}$, which is consistent with the result from \cite{2022ApJ...925..178H}. $f_{m_{1}>8} \cong 0.006$, and $f_{m_{1}>18} \cong 0.002$ are the fraction of stars with masses greater than $8\,M_\odot$ or $18\,M_\odot$, based on the initial mass function from \cite{2001MNRAS.322..231K}. $f_{\rm triple}$ and $f_{\rm binary}$ represent the binary and triple fractions for massive stars, respectively. The probability of a triple is treated as a function of the primary mass $M_1$, based on the observational results of \cite{2017ApJS..230...15M}. This probability depends on $M_1$ as well as the likelihood of its companions forming at different orbital periods. In other words, $f_{\rm triple}$ and $f_{\rm binary}$ are obtained by convolving the initial mass function from \cite{2001MNRAS.322..231K} with the triple fraction, similar to the approach used in the study by \cite{2023ASPC..534..275O}. Using this method, we obtain $f_{\rm triple} = 84\%$ and $f_{\rm binary} = 13\%$. It is worth noting that these results are based on extrapolation down to a minimum mass ratio of $q_{\rm min} = 0.01$. All of the above fractions are independent of the outcomes from our models. Only the overall fraction of system that evolve into dBH-LCs, $f_{\rm dBH-LC}$, depends on the outcome of our population synthesis. It is important to emphasize that $f_{\rm dBH-LC}$ is inevitably a function of the assumed initial mass and separation distributions, making it sensitive to model uncertainties.

\section{Results}
Using MOBSE and TSE, we evolve the initial massive binary and triple populations sampled in Section 2.3. In the following sections, we first present two examples of dBH-LCs formed through triple evolution. Then, we show a series of results for dBH-LCs formed through IBE and triple evolution.

\subsection{Example channels for dBH-LC formation from triple systems}
In the simulations of triple evolution, dBH-LCs are primarily formed through two channels:

1. The inner binary undergoes a merger, and the merger product later evolves into a BH. In this case, it forms a dBH-LC with the original tertiary.

2. One of the massive stars in the triple undergoes a SN. If the inner orbit remains bound after the SN but the outer orbit is disrupted, the remaining binary may evolve into a dBH-LC at a later stage.

\begin{figure*}[htbp]
\centering
\includegraphics[height=9cm, keepaspectratio]{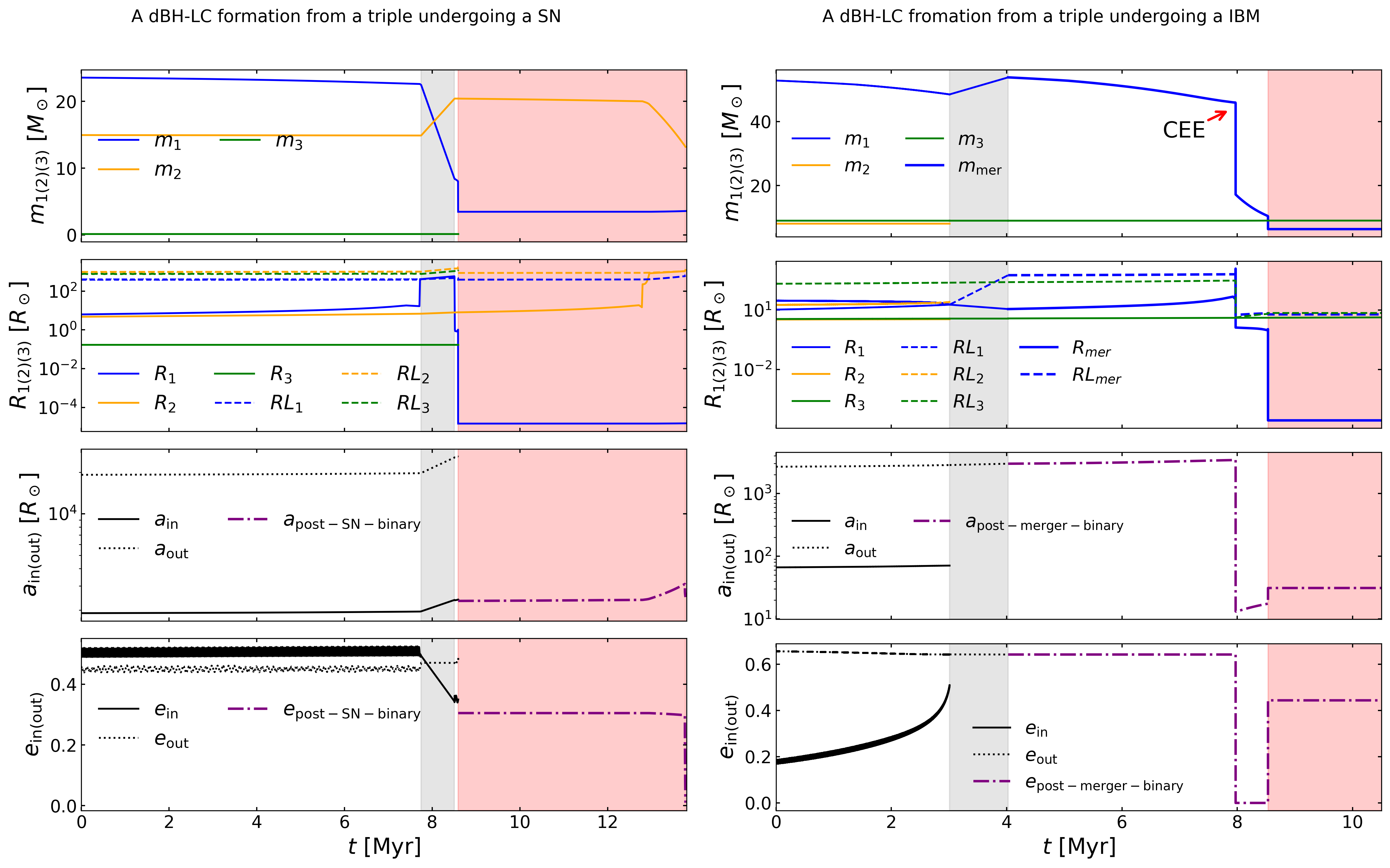}
\caption{Examples of dBH-LC formation through SN or inner binary merger in triple evolution are shown. Both panels display the evolution of component mass, radius, orbital separation, and eccentricity as functions of time. In both panels, the gray region indicates the phase when the inner binary undergoes RLOF, and the red region marks the dBH-LC phase. In addition, the initial parameters of the two triples shown in the figure are as follows:
In the left panel, the masses of the system are $m_{\rm 1}$ = 23.57 $M_{\odot}$, $m_{\rm 2}$ = 14.96 $M_{\odot}$, and $m_{\rm 3}$ = 0.13 $M_{\odot}$. The semi-major axes and eccentricities are $a_{\rm in}$ = 1902.10 $R_{\odot}$, $a_{\rm out}$ = 19130.05 $R_{\odot}$, $e_{\rm in}$ = 0.51, and $e_{\rm out}$ = 0.46. The inclinations and arguments of pericenter are $\cos i_{\rm in}$ = 0.80, $\cos i_{\rm out}$ = 0.04, $\omega_{\rm in}$ = 6.00, and $\omega_{\rm out}$ = 0.07. In the right panel, the masses of the triple are $m_{\rm 1}$ = 52.84 $M_{\odot}$, $m_{\rm 2}$ = 8.07 $M_{\odot}$, and $m_{\rm 3}$ = 9.00 $M_{\odot}$. The semi-major axes and eccentricities are $a_{\rm in}$ = 66.26 $R_{\odot}$, $a_{\rm out}$ = 2666.19 $R_{\odot}$, $e_{\rm in}$ = 0.18, and $e_{\rm out}$ = 0.66. The inclinations and arguments of pericenter are $\cos i_{\rm in}$ = -0.98, $\cos i_{\rm out}$ = -0.99, $\omega_{\rm in}$ = 0.79, and $\omega_{\rm out}$ = 2.99.}
\label{fig:2}
\end{figure*}

In Fig. \ref{fig:2}, we show two typical examples of dBH-LC formation. In the left panel, the triple experiences strong ZLK oscillations. Because the inner orbit is wide (about 1900 $R_{\odot}$), the tidal effect is weak. As the primary evolves, it begins RLOF at periastron at $\sim$ 7.75 Myr. During the non-conservative MT process (gray area), the donor's mass decreases from 22.5 $M_{\odot}$ to 8.4 $M_{\odot}$. The accretor's mass increases from 14.9 $M_{\odot}$ to 20.4 $M_{\odot}$. About 8.6 $M_{\odot}$ of mass is lost from the binary system. The inner eccentricity is reduced from 0.49 to 0.34. In the later evolution, the primary becomes a helium star. At 8.6 Myr, it undergoes a SN and forms a BH with a mass of 3.5 $M_{\odot}$. After the SN, the outer orbit is disrupted, but the inner binary stays bound. At this point, a dBH-LC system is formed (red area), with an eccentricity of 0.3, a separation (orbital period) of 2332 $R_{\odot}$ (2671 days), and masses of 3.5 $M_{\odot}$ for BH and 20.4 $M_{\odot}$ for the LC.

In the right panel of Fig. \ref{fig:2}, the inner binary merger (IBM) during the evolution of the triple. Later, the outer binary evolves and forms a dBH-LC. During the evolution, because the initial inner orbital separation is small (about 66 $R_{\odot}$), the ZLK effect of the triple is strongly suppressed by tides, and the periodic oscillation of the inner eccentricity ($e_{\rm in}$) is very weak. The inverse of the spin periods ($\frac{1}{\Omega}$) of the two stars in the inner binary is very short ($\sim$ 0.43 days for the primary and $\sim$ 0.13 days for the secondary). This is much shorter than the orbital period of the inner binary ($\sim$ 8 days). This causes the rotation angular momentum of the two stars in the inner binary to transfer to the inner orbit, which causes $e_{\rm in}$ and $a_{\rm in}$ to increase (A similar conclusion was also found in the analysis by \cite{2011CeMDA.111..105C}, \cite{2016CeMDA.126..189C}, \cite{2019A&A...627A.109Z}, and the analysis in Section 2.1.2 of \cite{2022MNRAS.516.1406S}). At about 3.01 Myr, the primary undergoes RLOF while on the MS stage. Because the mass ratio is too small (q = $\frac{8.1}{48.5}$ = 0.167), a merger occurs. According to the assumption by \cite{2002MNRAS.329..897H}, when two MS stars merge, their material mixes completely, and the merger product remains a MS star. Later, at 7.97 Myr, the merger product underwent mass transfer (MT) with its companion (the original tertiary) during the HG phase. Because the mass ratio was too small (q = $\frac{9.0}{45.9}$ = 0.196), this MT led to a CEE. During the CEE, the post-merger binary survived, and the envelope of the primary was successfully ejected. It became a 17.2 $M_{\odot}$ WR star, and the orbital separation shrank from 3421 $R_{\odot}$ to 13 $R_{\odot}$. Due to strong stellar winds during the WR phase, it becomes a WR star of about 10.6 $M_{\odot}$ before undergoing SN. In the end, it forms a 6.4 $M_{\odot}$ BH through complete fallback, and together with a 9 $M_{\odot}$ LC (the original tertiary) to form a dBH-LC system (red area).

\subsection{Population synthesis results}

In this section, we first focus on the properties of all dBH-LCs, and then narrow our focus to those dBH-LCs that are potentially observable by Gaia. We divide the dBH-LC population into dBH-MS and dBH-post MS (dBH-PMS). dBH-MS are usually binaries that form just after the BH is born. For dBH-LCs formed through the IBE channel, we classify them into three sub-channels: No MT channel: None of the stars in the binary or triple undergo MT before forming the dBH-LC;
MT channel: The stars experience only stable MT before forming the dBH-LC; CE channel: At least one episode of common envelope evolution (CEE) occurs before the dBH-LC is formed. For triple evolution, as described in Section 3.1, we divide the formation channels into IBM, and SN.

\begin{figure*}[htbp]
\centering
\includegraphics[width=1\textwidth]{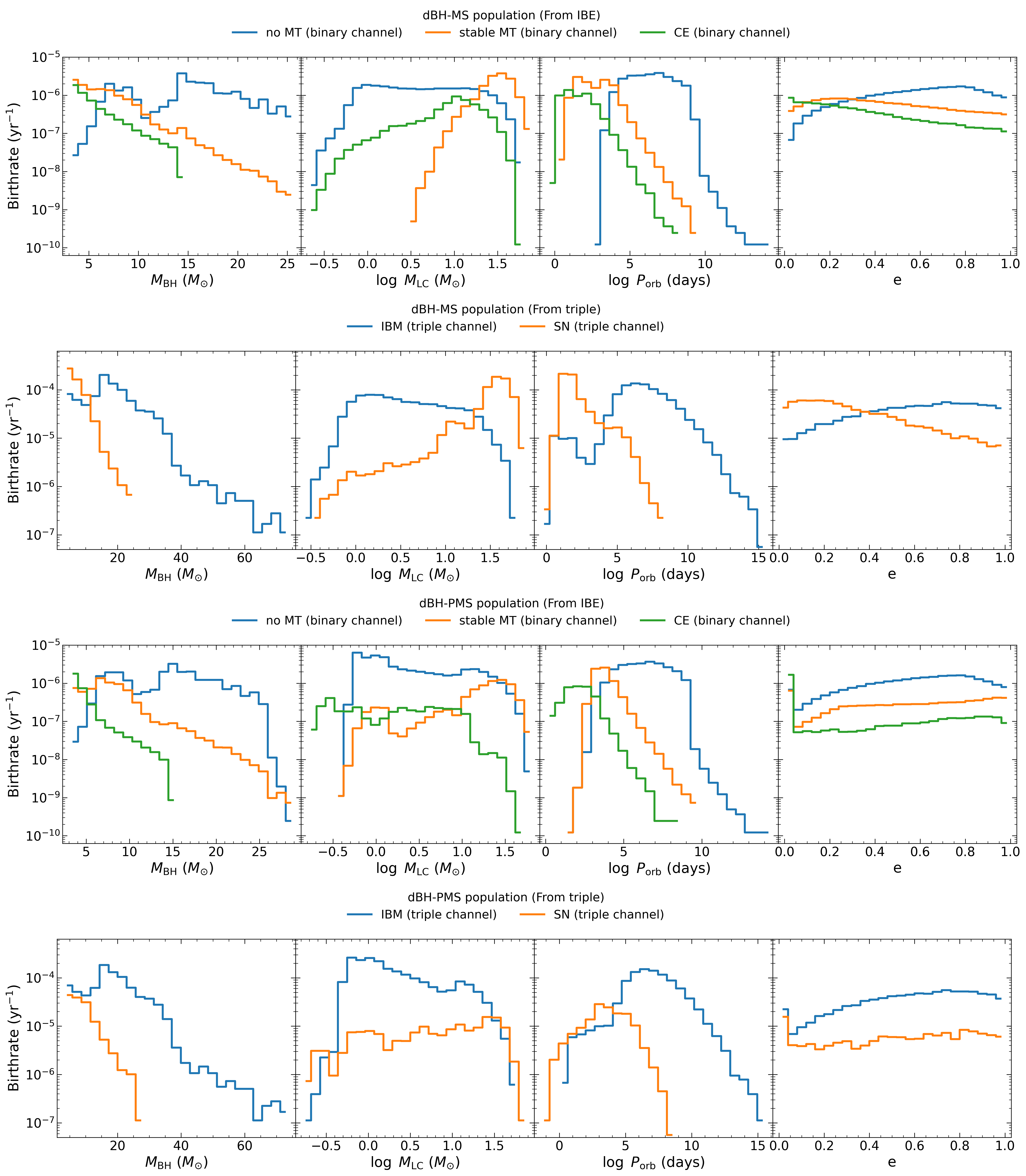}
\caption{At the metallicity of the MW, the birthrate distributions of the physical parameters of dBH-MS and dBH-PMS formed through IBE and triple evolution.}
\label{fig:3}
\end{figure*}

Fig. \ref{fig:3} shows the birthrate distributions of dBH-MS and dBH-PMS formed through IBE and triple evolution. In the Fig. \ref{fig:3}, the birthrates of the parameter distributions of dBH-MS formed from triple evolution are one to two orders of magnitude higher than those formed from IBE. For $M_{\rm BH}$, the maximum $M_{\rm BH}$ formed by IBE ($\sim$ 25 $M_{\odot}$) is about three times smaller than that formed by triple evolution (about 73 $M_{\odot}$, see also \cite{2024ApJ...964...83G}). In triple evolution, the $M_{\rm BH}$ ($>$ 25 $M_{\odot}$) come from the IBM channel (see the analysis below for details). The $M_{\rm BH}$ formed by the SN channel and the IBM channel both show an approximately monotonic decrease. For IBE, low-mass BHs ($<$ 5 $M_{\odot}$) mainly come from the CE channel. These BHs mainly form through accretion-induced collapse and core-collapse SN. When BHs with masses greater than 15 $M_{\odot}$ form, the larger the $M_{\rm BH}$, the stronger the fallback \citep{2012ApJ...749...91F}. When BHs form through full fallback, they experience very weak kicks. This increases the birthrate of dBH-MS with BHs larger than 15 $M_{\odot}$. In addition, it allows the formation of dBH-MS with BHs larger than 15 $M_{\odot}$ in the long orbital period range ($>$$10^{4.0}$ days). For $M_{\rm LCs}$, both triple evolution and IBE tend to form more massive LCs ($>$ 10 $M_{\odot}$). This is because more massive LC survive more easily in the CEE process, or they are more likely to remain bound in the SN process due to higher gravitational binding energy. For $P_{\rm orb}$ and eccentricity ($e$), the dBH-MS formed by triple evolution and IBE both show traces of the initial outer binary distribution and the initial IBE distribution.

\begin{figure*}[htbp]
\centering
\includegraphics[width=1\textwidth]{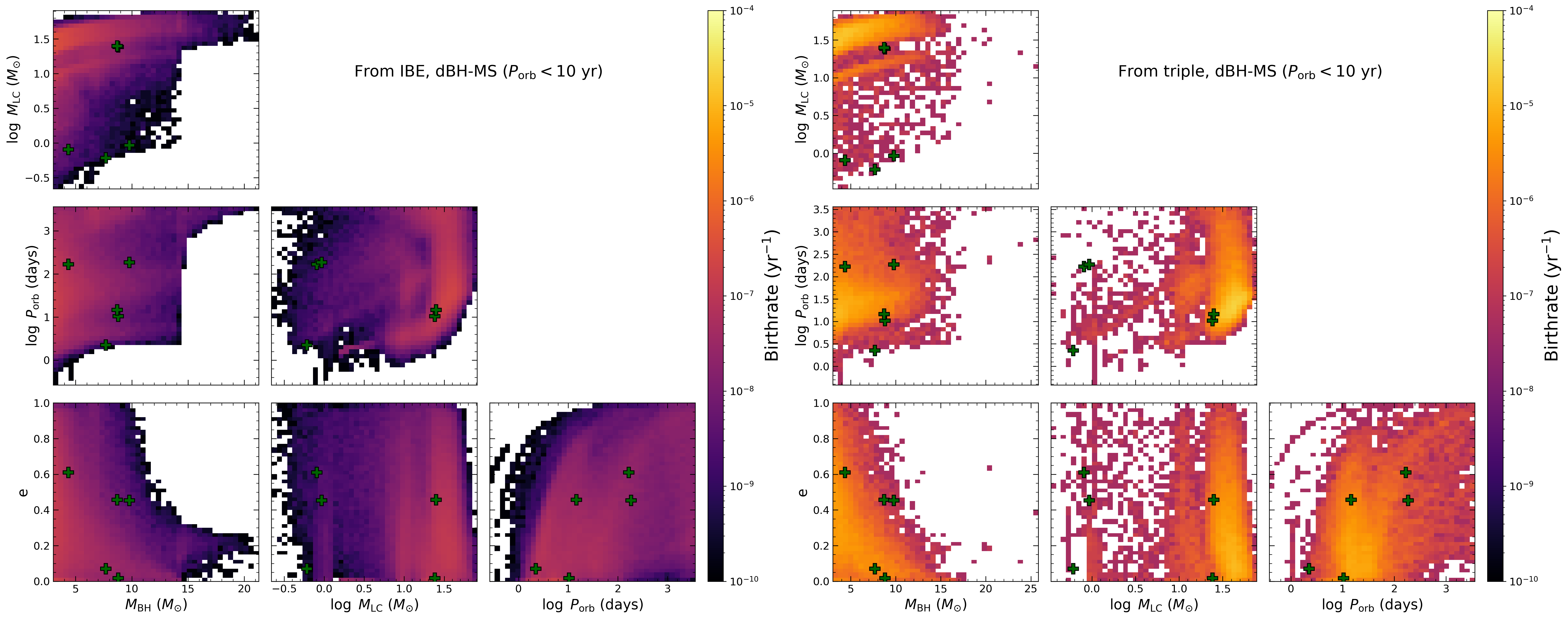}
\caption{The birthrate 2D distribution of dBH-MS in the MW is shown as a function of LC mass ($M_{\rm LC}$), BH mass ($M_{\rm BH}$), orbital period ($P_{\rm orb}$), and eccentricity ($e$). The color in each pixel is scaled according to the birthrate of dBH-LC binaries. The left panels show dBH-MS formed through IBE, and the right panels show those formed through triple evolution. The darkgreen crosses represent the observational data from Table 1. It is worth noting that our study focuses more on the differences between the dBH-LC populations formed through IBE and triple evolution. Therefore, in Table 1, we classify the observations only by the stellar types of the LCs. We emphasize that explaining these observations still requires attention to additional physical properties (for example, the metallicity of each observation).}
\label{fig:4}
\end{figure*}

\begin{figure*}[htbp]
\centering
\includegraphics[width=1\textwidth]{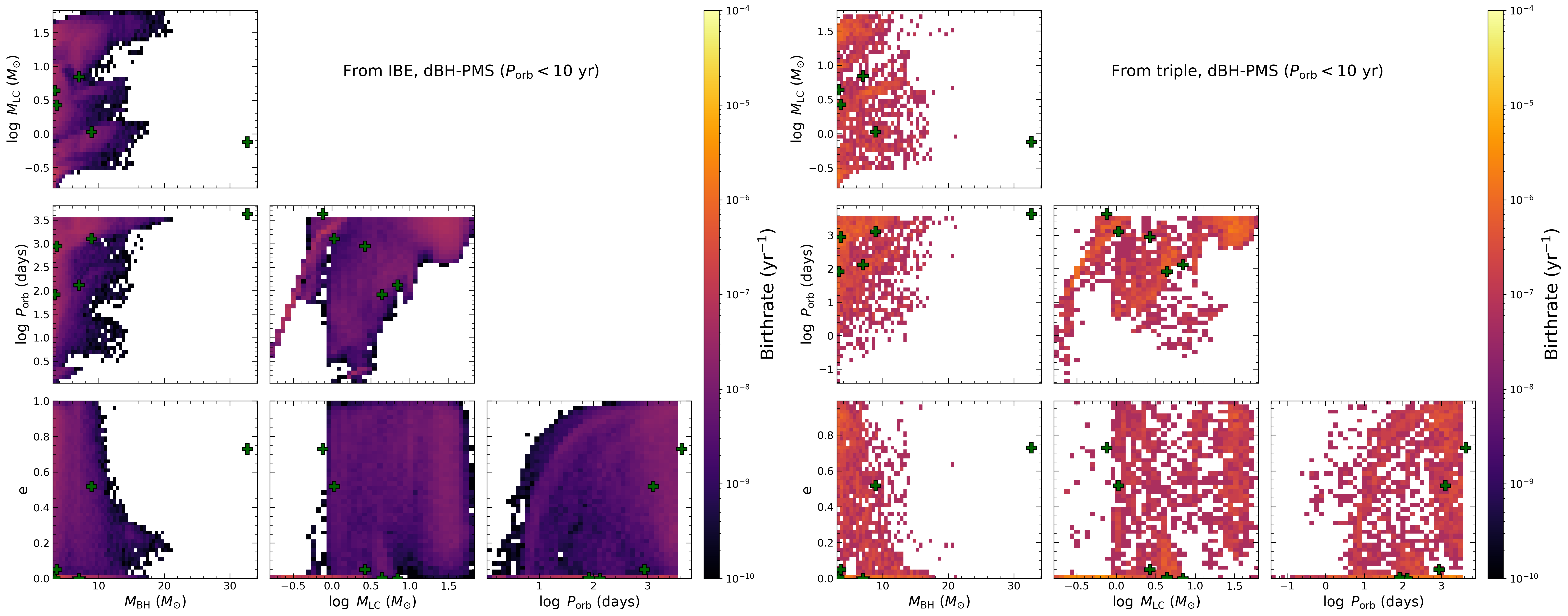}
\caption{Similar to Fig. \ref{fig:4}, but for dBH-PMS.}
\label{fig:5}
\end{figure*}

Considering that $Gaia$ has a maximum observation duration of 10 ${\rm yr}$ for dBH-LCs, we focus on dBH-LCs with $P_{\rm orb} < 10\ {\rm yr}$. Fig. \ref{fig:4} and Fig. \ref{fig:5} present the 2D histograms of dBH-MS and dBH-PMS with $P_{\rm orb} < 10\ {\rm yr}$, formed through IBE and triple evolution. The green crosses indicate the observational data from Table 1. In Fig. \ref{fig:4} and Fig. \ref{fig:5}, we find that both IBE and triple evolution can basically explain the pairwise physical quantities of these observations. However, our model has difficulty explaining the $M_{\rm BH}$ of Gaia BH3. This is model dependent and may vary significantly with the assumed metallicity, as Gaia BH3 is extremely metal poor. The birthrate of dBH-LCs in each bin from triple evolution is generally higher than that from the IBE model. This means that triple evolution has a greater advantage in explaining the currently observed dBH-LCs. However, we note that our triple evolution model is still insufficient for the statistical properties of dBH-LCs with $P_{\rm orb} < 10\ {\rm yr}$. This is because the outer orbital periods in our initial sampling of triples are mainly concentrated around $10^{5}$ ${\rm days}$. In triple evolution, the formation of dBH-LCs is mainly contributed by the IBM channel, and most of the dBH-LCs formed through this channel have $P_{\rm orb} < 10\ {\rm yr}$ (see also Fig. \ref{fig:3}). In addition, by comparing Fig. \ref{fig:4} and Fig. \ref{fig:5}, we find that some PMS stars may interact with the BH. This causes some low-mass BH to merge with the PMS in the CEE process. It also causes the stable MT channel and the CE channel in dBH-PMS to produce more low-mass LC ($<$1\ $M_{\odot}$) and circularized dBH-PMS systems ($e$ = 0). In addition, the mass-loss rate of PMS stars is usually higher than that of MS stars. This stronger mass-loss rate can drive orbital expansion. This causes the orbital periods of dBH-PMS to be generally higher than those of dBH-MS, such as Gaia BH1/2/3.

Table 2 shows the birthrates of dBH-LCs formed by IBE and triple evolution in the MW. In the IBE model, the birthrate of dBH-LCs is 4.35$\times$$10^{-5}$ ${\rm yr}^{-1}$. This is similar to the result calculated by \cite{2019ApJ...885..151S} (see Table 1 of \cite{2019ApJ...885..151S}). In the triple evolution model, the birthrate of dBH-LCs is 1.47$\times$$10^{-3}$ ${\rm yr}^{-1}$. This is one or two orders of magnitude higher than the birthrate of dBH-LCs formed by IBE, which can be explained by two key factors: the higher triple fraction of massive stars (contributing at most one order of magnitude), and more importantly, the effect of triple dynamics, which significantly enhances the birthrate of dBH-LCs through triple evolution. This result shows that the evolution of massive stars in triples is more likely to form dBH-LCs. This conclusion is similar to that of \cite{2024ApJ...964...83G}. In addition, an LC in the PMS phase must have evolved through the MS phase. However, not all dBH-MS systems will later evolve into dBH-PMS systems. This is because, during PMS evolution, the LC in a dBH-MS system may interact with the BH — such as through CEE. These interactions may result in the formation of an BH X-ray binary, or in a merger between the LC and BH (e.g., if the system does not survive CEE). As a result, in both the IBE model and the triple evolution model, the birthrate of dBH-MS is higher than that of dBH-PMS. Furthermore, estimating the number of dBH-LCs is timely for the upcoming Gaia DR4. In detail, we estimate the number by multiplying the birthrate with the average lifetime of LCs in different mass bins. Because the recently observed dBH-LCs are not far from us (for example, Gaia BH1/2/3 are at about 480 pc \citep{2023MNRAS.518.1057E}, 1.16 kpc \citep{2023MNRAS.521.4323E}, and 590 pc \citep{2024A&A...686L...2G} from the Sun), we calculate the birthrate and number of nearby dBH-LCs, as shown in Table 3. In this calculation, we assume that dBH-LC remain close to their birth locations and do not consider their possible motion. Our sampling method for the distance between a dBH-LC and the Sun is similar to Section 2.4 of \cite{2024MNRAS.529.1886H}. There are still few studies that estimate the total number of dBH-LCs through triple evolution. Most previous works focus on isolated binary evolution or dynamical evolution. \cite{2017ApJ...850L..13B}, \cite{2022ApJ...931..107C}, \cite{2019ApJ...885..151S}, \cite{2024ApJ...965...22D} and \cite{2025PASP..137d4202N} predicted the number of observable dBH-LCs as 3800$-$12000, 30$-$300, 470$-$12000, $\sim$155724 and $\sim$19466, respectively. These differences can be explained by different model assumptions and choices of observational effects. In Table 3, our binary evolution model predicts about 3000 dBH-LCs. This number is lower than most of their predictions. The difference comes from our lower initial binary fraction (13\% in our work). Finally, by combining the binary evolution channel and the triple evolution channel, we estimate that the entire MW contains approximately 87,000 dBH-LCs with $P_{\rm orb} < 10$ yr. Among them, about 75,000 dBH-LCs are within 20 kpc from the Sun, and about 570 dBH-LCs are within 1 kpc.

\begin{table}[ht]
\centering
\caption{The birthrates of dBH-LCs (including dBH-MS and dBH-PMS), dBH-MS, and dBH-PMS under different evolution models at the metallicity of the MW.}
\label{tab:2}

\begin{threeparttable}

\begin{tabular}{lccc}
\hline\hline
Model & $R_{\rm dBH\mbox{-}LC}$ (yr$^{-1}$) & $R_{\rm dBH\mbox{-}MS}$ (yr$^{-1}$) & $R_{\rm dBH\mbox{-}PMS}$ (yr$^{-1}$) \\
\hline
Binary & $4.35\times10^{-5}$ & $4.26\times10^{-5}$ & $3.10\times10^{-5}$ \\
Triple & $1.47\times10^{-3}$ & $1.43\times10^{-3}$ & $9.42\times10^{-4}$ \\
\hline
\end{tabular}

\begin{tablenotes}
\footnotesize
\item \textbf{Notes.}
The first column shows the different evolutionary models. The second column shows the metallicity. The third, fourth, and fifth columns show the birthrates of dBH-LC, dBH-MS, and dBH-PMS, respectively.
\end{tablenotes}

\end{threeparttable}
\end{table}

\begin{table*}[ht]
\centering
\caption{The birthrate and number (in parentheses) of dBH-LCs with orbital period $P_{\rm orb} < 10$ yr.}
\label{tab:3}

\begin{threeparttable}

\begin{tabular}{llccc}
\toprule
Model & Type & Birthrate (No.) (All) & Birthrate (No.) ($d<20\ \mathrm{kpc}$) & Birthrate (No.) ($d<1\ \mathrm{kpc}$) \\
\midrule
\multirow{2}{*}{Binary}
  & dBH-MS  & $1.36\times10^{-5}\ \mathrm{yr}^{-1}\ (3318)$ & $1.17\times10^{-5}\ \mathrm{yr}^{-1}\ (2853)$ & $9.00\times10^{-8}\ \mathrm{yr}^{-1}\ (22)$ \\
  & dBH-PMS & $4.08\times10^{-6}\ \mathrm{yr}^{-1}\ (224)$  & $3.51\times10^{-6}\ \mathrm{yr}^{-1}\ (193)$  & $2.70\times10^{-8}\ \mathrm{yr}^{-1}\ (1)$ \\
\addlinespace
\multirow{2}{*}{Triple}
  & dBH-MS  & $5.22\times10^{-4}\ \mathrm{yr}^{-1}\ (76734)$ & $4.49\times10^{-4}\ \mathrm{yr}^{-1}\ (65991)$ & $3.46\times10^{-6}\ \mathrm{yr}^{-1}\ (508)$ \\
  & dBH-PMS & $6.91\times10^{-5}\ \mathrm{yr}^{-1}\ (6910)$  & $5.94\times10^{-5}\ \mathrm{yr}^{-1}\ (5943)$  & $4.57\times10^{-7}\ \mathrm{yr}^{-1}\ (46)$ \\
\bottomrule
\end{tabular}

\begin{tablenotes}
\footnotesize
\item \textbf{Notes.}
The first column shows the different evolution models. The second column shows the types of dBH-LCs. The third, fourth, and fifth columns show the birthrate and number for different distances with $P_{\rm orb} < 10$ yr.
\end{tablenotes}

\end{threeparttable}
\end{table*}

\begin{figure*}[htbp]
\centering
\includegraphics[width=0.8\textwidth,height=0.9\textheight]{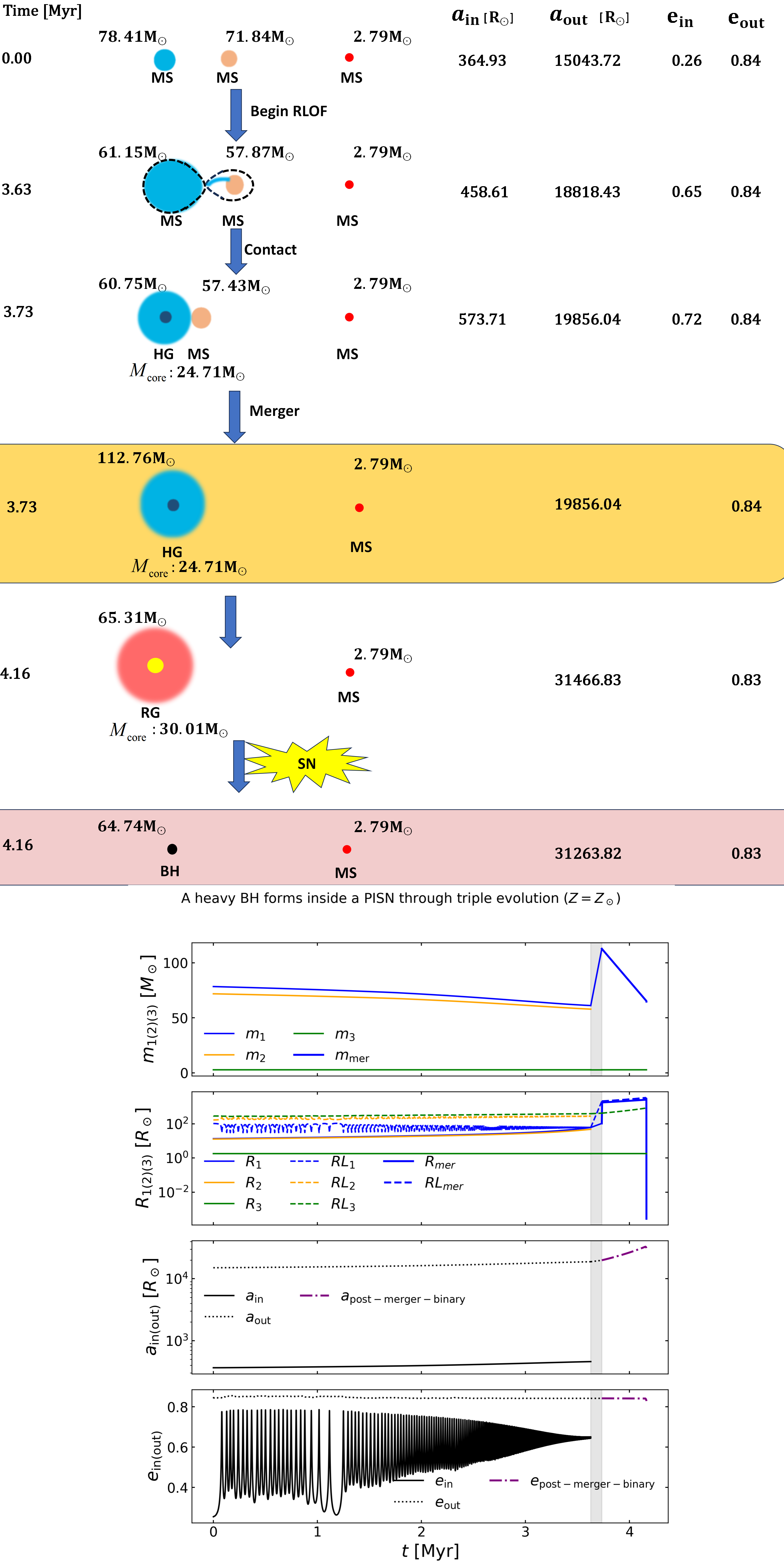}
\caption{An example of a dBH-LC with a black hole in the PISN range formed through triple evolution. In the upper panel, the yellow area and the red area represent the formation of the post-merger binary and the formation of the dBH-LC, respectively. In the lower panel, the gray area represents the phase when the inner binary undergoes mass transfer.}
\label{fig:6}
\end{figure*}

In addition, we note that triple evolution can produce some dBH-LCs with more massive BHs that cannot be formed through IBE (e.g., $M_{\rm BH} > 30\ M_{\odot}$, see also Fig. \ref{fig:3}). As analyzed for Fig. \ref{fig:3}, these dBH-LCs with more massive BHs are formed through the IBM channel of triple evolution. In Fig. \ref{fig:6}, we show an example of how a dBH-LC with a massive BH ($>$ 60\ $M_{\odot}$) can form through triple evolution. In the triple evolution of Fig. \ref{fig:6}, a primary with an initial mass of 78.41 $M_{\odot}$ evolves first. The inner binary then undergoes strong ZLK oscillations, and its inner eccentricity shows long-term periodic changes between 0.25 and 0.78. As the radius of the primary expands, tidal effects gradually increase. When the precession caused by tides dominates over the precession caused by the ZLK mechanism, the inner binary can effectively decouple from the tertiary \citep{2022MNRAS.516.1406S}. This eventually suppresses any ZLK oscillations, reducing the periodic oscillations of the inner eccentricity \citep{2016ComAC...3....6T,2024MNRAS.527.9782D}. The primary undergoes RLOF at about 3.63 Myr. As MT continues, the secondary in the inner binary also expands. At about 3.73 Myr, the primary evolves into a HG star and comes into contact with the secondary (an MS star), leading to a merger. According to the collision matrix of \cite{2002MNRAS.329..897H}, the 3D simulations of binary mergers by \cite{2013MNRAS.434.3497G}, \cite{2022MNRAS.516.1072C} and \cite{2022MNRAS.516.1072C}, and the simulations of binary mergers and the evolution of merger products using the 1D stellar evolution code MESA by \cite{2020ApJ...904L..13R}, \cite{2022MNRAS.516.1072C} and \cite{2022MNRAS.516.1072C}, the secondary (MS star) dissolves in the envelope of the HG star. This makes the merger product have a larger envelope mass, and finally forms a merger product with a small helium core mass (about 24.71 $M_{\odot}$) but a large envelope mass (about 88.05 $M_{\odot}$). This merger product undergoes significant mass loss in the later evolution. However, before the SN, it still has about 35.3 $M_{\odot}$ of envelope mass. This causes a large fallback during the SN, with a negligible natal kick. Finally, it forms a dBH-LC composed of a 64.74 $M_{\odot}$ BH and a 2.79 $M_{\odot}$ MS star (the original tertiary). Therefore, through triple evolution, BH binaries where BH mass lies in PISN range (about $60\ M_{\odot} \sim 120\ M_{\odot}$) can be produced. We also calculate that the birthrate of dBH-LCs with a BH in the PISN range, formed in this way at thin disk metallicity, is about 1.78$\times$$10^{-6}$ ${\rm yr}^{-1}$.

\begin{figure}[htbp]
\centering
\includegraphics[width=0.5\textwidth]{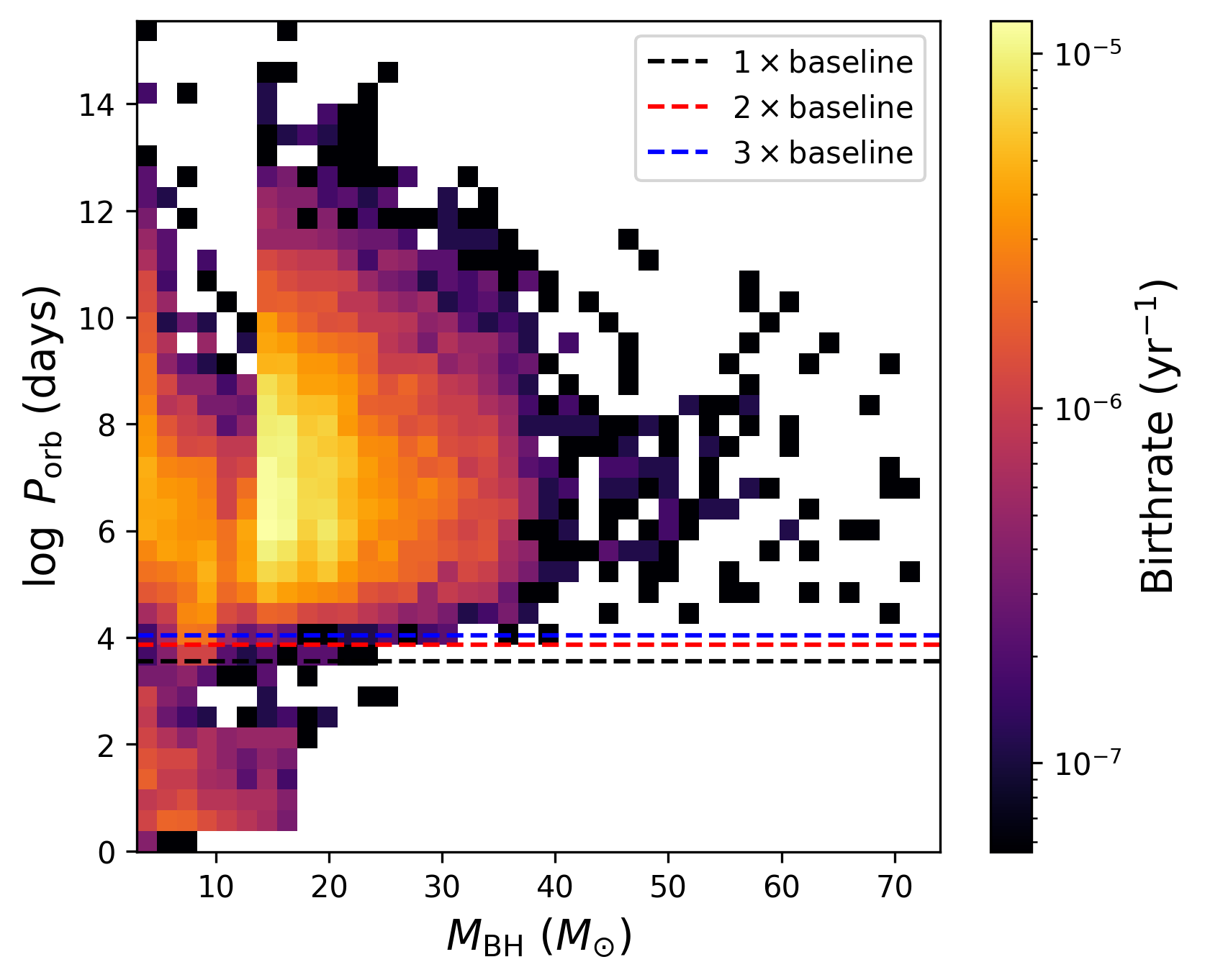}
\caption{The 2D birthrate distribution of $M_{\rm BH}$ $-$ $P_{\rm orb}$ for dBH-MS formed through the IBM channel in triple evolution. This also shows 1$\times$, 2$\times$, and 3$\times$ Gaia’s maximum observational baseline (10 years), which correspond to the cases where Gaia can at most fully resolve the entire orbit, half of the orbit, and one-third of the orbit, respectively.}
\label{fig:7}
\end{figure}

In Fig. \ref{fig:7}, we present the 2D histogram of dBH-LC systems formed through the IBM channel in triple evolution. In Fig. \ref{fig:7}, dBH-LCs with $M_{\rm BH}$ greater than 30 $M_{\odot}$ have generally very long orbital periods ($P_{\rm orb} > 10^{4}\ {\rm days}$), which are much longer than the observation duration of Gaia. Compared to Gaia’s maximum observational baseline (10 years), the maximum BH masses for which Gaia can at most fully resolve the orbit, resolve half of the orbit, and resolve one-third of the orbit are approximately 25 M${_\odot}$, 30 M${_\odot}$, and 40 M${_\odot}$, respectively. This is because the merger products of the progenitor stars that form these massive BHs have very high masses. Their strong stellar winds can expand the orbital periods of the post-merger binaries to very large. In addition, dBH-LC systems formed through the IBM channel show a gap in orbital periods between $\sim$ $10^2$ and $10^3$ days. The main reason for this gap is that, in the IBM channel, the merger product and its LC either undergo CEE and form a short-period system, or they never interact and form a long-period system. This result is very similar to the findings of \cite{2025PASP..137d4202N} (see their Figure 1).

\section{Discussion}
As discussed in Section 3, although we have considered the formation of dBH-LC populations under different evolutionary models, metallicities, and formation channels, the evolution of massive stars is highly complex. Therefore, our conclusions still carry uncertainties. As in the recent work by \cite{2025A&A...698A.240S}, our study is based on the rapid stellar evolution (RSE) code that uses the fitting formulas from \cite{2000MNRAS.315..543H,2002MNRAS.329..897H}. However, this approach still shows significant differences from detailed stellar evolution codes (e.g., MESA code \citep{2011ApJS..192....3P,2013ApJS..208....4P,2015ApJS..220...15P,2018ApJS..234...34P,2019ApJS..243...10P}) in modeling the evolution of massive stars. Typically, these fitting formulas do not self-consistently account for the effects of mass loss on stellar evolution \citep{2023NatAs...7.1090B}. As pointed out in the works of \cite{2025A&A...698A.240S} and \cite{2025ApJ...983..115S}, compared to the MESA code, the SSE code tends to overestimate the radii of massive stars and their stellar winds. These differences can lead to significant discrepancies in the predicted final outcomes and orbital evolution of massive stars. For example, RSE code may result in an overestimation of the interaction frequency among massive stars, affect the efficiency of tidal interactions, and underestimate the effectiveness of the ZLK mechanism in triple evolution. In addition, the TSE code currently does not model systems undergoing dynamical instability or RLOF involving the tertiary. In the study by \cite{2021MNRAS.500.1921G}, \cite{2022ApJS..259...25H}, \cite{2022ApJ...925..178H}, \cite{2025ApJ...979L..37L}, \cite{2025A&A...693A..14B} and \cite{2025A&A...693A..84K} it was shown that MT from the tertiary to the inner binary or dynamical instability in triple can lead to inner binary mergers or the ejection of the lowest-mass star. The contribution of such scenarios to the dBH-LC population remains unknown.

Besides, for BH SN kicks, we adopt the more recent prescription proposed by \cite{2020ApJ...891..141G}. This SN model can explain a wide range of observational constraints, including the merger rates of double neutron stars (BNS), BBH, and NS-BH binaries inferred by  LIGO-Virgo collaboration (e.g., GW170817) \citep{2019ApJ...882L..24A}, the natal kicks of young pulsars, and the differences in kick between core-collapse SNe, electron-capture SNe, and ultra-stripped SNe in binary systems. Specifically, this SN prescription accounts for the remnant mass, the fallback fraction, and the ejecta mass (see Equation 4). According to this model, the natal kick distribution of BHs shows two peaks: one at $\sim$ 0 km/s (corresponding to BHs formed via direct collapse, which make up about 60\% of the population), and another at tens of km/s ($\le$ 100 km/s). The average natal kick velocity for BHs under this prescription is $\sim$ 30 km/s (see Table 2 in \cite{2020ApJ...891..141G}). At present, the distribution of BH natal kicks is still not well understood. Recent studies support the idea that many BHs receive negligible natal kicks at birth \citep{2003Sci...300.1119M,2022NatAs...6.1085S,2024Natur.635..316B,2024PhRvL.132s1403V,2025ApJ...979..209V,2025PASP..137c4203N,2025ApJ...983..115S} — in particular, \cite{2024Natur.635..316B} and \cite{2025ApJ...983..115S} found that known BHs in triple likely formed without any significant kicks. However, some BHs do appear to receive relatively large natal kicks, ranging from tens to over one hundred km/s \citep{2022ApJ...930..159A,2023ApJ...952L..34K,2025A&A...693A.129M}. In addition, previous studies have shown that low natal kicks ($<$ 40$\sim$50 km/s) are generally favored for the formation of wide-orbit BH binaries (e.g., Gaia BH1/2/3 \citep{2023MNRAS.518.1057E,2023MNRAS.521.4323E,2024MNRAS.535.3577K,2024ApJ...975L...8L}).

Due to observational limitations, the statistics of binaries or multiple systems with low mass ratios remain highly uncertain, mainly because of the large brightness contrast. In this work, we extrapolate the minimum companion mass down to the brown dwarf boundary, assuming a hydrogen-burning limit of 0.08 M$_{\odot}$. Our extrapolated results are consistent with those of \cite{2024ApJ...964...83G}, who also found that lowering the minimum mass ratio leads to a significantly increased probability of forming companions at wider separations (with a typical peak orbital period around $10^{5.5}$ days; see also Fig. 1). This suggests that primary is very likely to host one or more companions in wide orbits (i.e., to form in binary or multiple systems), but such companions are extremely difficult to detect observationally, which introduces large uncertainties in the statistics of low mass-ratio companions. On the other hand, our study is limited to systems with up to triple. In reality, the probability of forming higher-order multiples increases significantly with the mass of the primary (see also Figure 39 in \cite{2017ApJS..230...15M}). This implies that the binary and triple fractions used in our convolution also carry uncertainties.

Additionally, this study is limited by the assumption of a single metallicity value (Z $=$ $Z_{\odot}$ $=$ 0.014) to represent that of the MW. In reality, the thick disk of the MW is characterized by lower metallicity, typically Z $=$ 0.15 $Z_{\odot}$ \citep{2013pss5.book..393Y}. It is well known that as metallicity decreases, the opacity of massive stars is reduced, leading to lower mass-loss rates and smaller maximum radii \citep{2001A&A...369..574V}. Consequently, orbital expansion in binary or triple is expected to be suppressed for low metallicity populations, and the probability of binary interactions is reduced. More importantly, massive stars with lower metallicity tend to form more massive BH, which is crucial for predicting dBH-LC with heavy BH (e.g.,Gaia BH3). Additionally, the reduced mass-loss rate in low metallicity environments allows massive stars to retain more of their rotational angular momentum during evolution. This allows rotationally induced mixing to drive chemically homogeneous evolution (CHE) in massive stars \citep{1975A&A....41..329Z,1981A&A....99..126H,1981A&A...102...17P,1987A&A...178..159M,1991A&A...241..419P,1992ApJ...391..246P,2005A&A...435.1013P,2005A&A...443..643Y,2006A&A...460..199Y,2009A&A...497..243D,2011A&A...528A.114T,2011A&A...530A.115B,2013A&A...552A.105V,2014ApJ...796...37S,2015A&A...573A..71K,2015A&A...581A..15S,2021ApJ...923..277R,2023ApJ...952...79L}. CHE is a key pathway for forming Wolf-Rayet stars \citep{2018PASP..130h4202C,2023A&A...674A.216L,2024ApJ...969..160L,2025RAA....25c5002H}, BHs \citep{2016MNRAS.458.2634M,2016A&A...588A..50M,2016MNRAS.460.3545D,2020A&A...641A..86H,2020MNRAS.499.5941D,2021MNRAS.505..663R,2025arXiv251008231L}, magnetars, and the progenitors of long gamma-ray bursts \citep{2004ApJ...607L..17P,2005A&A...443..643Y,2006A&A...460..199Y,2006ApJ...637..914W,2013A&A...554A..23M}. In particular, \cite{2024MNRAS.527.9782D} showed that in triple, the inner binary can undergo CHE and produce double BHs, which may eventually merge due to ZLK mechanism within the triple. This channel may also be important for the formation of dBH-LC.

\section{Conclusions}
In this study, we use the MOBSE and TSE codes to simulate the formation and evolution of dBH-LCs in the MW. We consider progenitor systems originating from both isolated binaries and hierarchical triples. In the MW, the birthrates of dBH-LC systems calculated from the IBE and triple evolution channels are 4.35$\times$$10^{-5}$ ${\rm yr}^{-1}$ and 1.47$\times$$10^{-3}$ ${\rm yr}^{-1}$, respectively. Compared to isolated binaries, massive stars in triples are more likely to interact within the inner binary due to ZLK oscillations. As a result, dBH-LC systems are predominantly formed through the channel in which the inner binary merges and forms a new binary (see Figs. \ref{fig:3}). In our calculation, the birthrate of dBH-LC from triple evolution is one to two orders of magnitude higher than that from the IBE channel. In addition, we find that the dBH-LC formed from post-merger binaries in triple systems have much larger $M_{\rm BH}$ than those formed through the IBE channel. The reason is that the merger product of a MS + CHeB inner binary is a massive star with a small helium core and a large envelope. Because the helium core is small and the total mass is large, these merger products can form BHs in the predicted PISN range (60 $M_{\odot}$ $\sim$ 120 $M_{\odot}$).

Below, we summarize the main results on the orbital properties of dBH-MS and dBH-PMS formed in the MW:

1. For dBH-MS, the $M_{\rm LC}$ from the IBE channel shows two peaks in the range of 10 $M_{\odot}$ to 63 $M_{\odot}$. The low-mass LCs are mainly contributed by the no-MT channel. At the same time, the eccentricities of dBH-MS from the IBE channel cover a wide range. The circularized dBH-MS are mainly contributed by the CE and stable MT channels. On the other hand, the $M_{\rm LC}$ of dBH-MS formed through triple evolution shows peaks at 1 $M_{\odot}$ and 63 $M_{\odot}$, mainly contributed by the IBM and SN channels, respectively.

2. For dBH-PMS, because some LCs undergo MT with the BH during later evolution, both the IBE and triple evolution channels produce more low-mass companions ($M_{\rm LC}$ $<$ 1.6 $M_{\odot}$) and circularized systems compared to dBH-MS.

\begin{acknowledgements}

This work received the support of the National Natural Science Foundation of China under grants U2031204, 12163005, 12373038, and 12288102; the Natural Science Foundation of Xinjiang No.2022TSYCLJ0006 and 2022D01D85; the China Manned Space Program grant No. CMS-CSST-2025-A15. AP23490322 — Exploration of Thermodynamic Properties of Relativistic Compact Objects Within the Framework of Geometrothermodynamics (GTD), Grant financing for scientific and/or scientific-technical projects for 2024–2026, Ministry of Science and Higher Education of the Republic of Kazakhstan.

\end{acknowledgements}
\bibliographystyle{aa}
\bibliography{aa55081-25}

\end{document}